\documentclass[aps,prl,twocolumn,superscriptaddress]{revtex4}
\usepackage{graphicx}
\usepackage{epstopdf}
\usepackage{color}

\begin{document}
\title{Unpaired Majorana modes in the gapped phase of Kitaev's honeycomb model}
\author{Olga Petrova}
\affiliation{Department of Physics and Astronomy, 
	The Johns Hopkins University,
	Baltimore, Maryland 21218, USA
}
\author{Paula Mellado}
\affiliation{School of Engineering and Applied Sciences, 
	Adolfo Ib{\'a}{\~n}ez University,
	Santiago, Chile
}
\author{Oleg Tchernyshyov}
\affiliation{Department of Physics and Astronomy, 
	The Johns Hopkins University,
	Baltimore, Maryland 21218, USA
}

\begin{abstract}

We study the gapped phase of Kitaev's honeycomb model (a $Z_2$ spin liquid) in the presence of lattice defects. We find that some dislocations and bond defects carry unpaired Majorana fermions. Physical excitations associated with these defects are (complex) fermion modes made out of two (real) Majorana fermions connected by a $Z_2$ gauge string. The quantum state of these modes is robust against local noise and can be changed by winding a $Z_2$ vortex around a dislocation. The exact solution respects gauge invariance and reveals a crucial role of the gauge field in the physics of Majorana modes.

\end{abstract}

\maketitle

In three dimensions all particles can be divided into two categories: bosons and fermions. In two dimensions, particles can exhibit statistics that interpolates continuously between Bose and Fermi's, hence the name \textit{anyons}. When two Abelian anyons are exchanged, the system's wavefunction picks up a phase that is not restricted to integer multiples of $\pi$ as it is in three dimensions. Non-Abelian statistics arises when the ground state of a system is degenerate and winding one particle around another amounts to a unitary transformation in the space of degenerate ground states. The nonlocal nature of the winding process makes the evolution of the quantum ground state insensitive to local noise. Therefore non-Abelian anyons could provide a pathway to fault-tolerant quantum computing. One of the main obstacles is the scarcity of systems with non-Abelian excitations. In this Letter, we discuss a scenario in which the addition of topological defects to a system with Abelian anyons can give rise to non-Abelian statistics. 

An exactly solvable spin model with Abelian anyon excitations was constructed by Kitaev \cite{Kitaev2006}. The model has one gapless and three gapped phases at zero temperature. The gapped phases have a $Z_2$ topological order, which yields Abelian anyonic statistics. Kitaev sketched a heuristic argument that a dislocation in his model should harbor an unpaired Majorana mode. Two such modes can be combined into a zero-energy fermion. The ground state of a system with a pair of distant dislocations is then doubly degenerate and can be manipulated by winding a vortex around one of the dislocations \cite{footnote}. In this Letter, we characterize the properties of dislocations in Kitaev's spin model and confirm his conjecture by an explicit construction of the Majorana modes. 

\begin{figure}
\centering
\includegraphics[width=0.99\columnwidth]{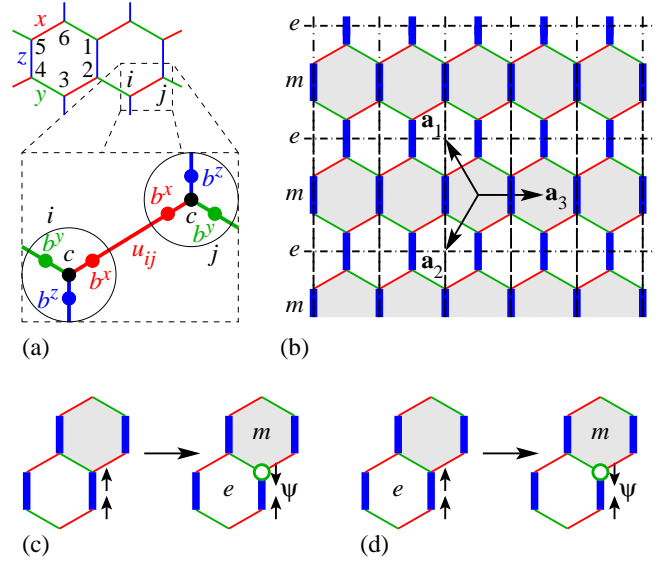}
\caption{(color) (a) Three types of bonds, the Majorana operators $b^\alpha$ and $c$ and the conserved plaquette flux $W = \sigma^{x}_{1}\sigma^{y}_{2}\sigma^{z}_{3}\sigma^{x}_{4}\sigma^{y}_{5}\sigma^{z}_{6}$. (b) Partitioning of honeycomb plaquettes into $e$ (blank) and $m$ (filled) in the gapped phase with strong $z$ links (thick lines). Dash-dotted lines are links of the toric-code square lattice. $\mathbf a_i$ are lattice vectors. (c)--(d) Local creation of an $e \times m$ vortex pair (c) and conversion of vortex flavor (d). Open green circle denotes the application of a $\sigma^y$ operator.}
\label{fig:intro}
\end{figure}

Kitaev's model has spins $1/2$ living on sites of a honeycomb lattice. They interact with their nearest neighbors through anisotropic exchange whose nature depends on the direction of the bond:
\begin{equation}
H = -J_x \sum_{x~\mathrm{links}} \sigma^{x}_{m}\sigma^{x}_{n}
	-J_y \sum_{y~\mathrm{links}} \sigma^{y}_{m}\sigma^{y}_{n}
	-J_z \sum_{z~\mathrm{links}} \sigma^{z}_{m}\sigma^{z}_{n},
\label{eq:HKitaev}
\end{equation}
where links are labeled as shown in Fig.~\ref{fig:intro}(a). Its solvability is related to the existence of one integral of motion for every hexagonal plaquette, $W = \prod\sigma^\mathrm{out}_{i}$, where ``out'' refers to the $x$, $y$, or $z$ label of the link pointing out of the plaquette on a given site, see Fig.~\ref{fig:intro}(a). In the ground state, all $W = +1$. The exact solution is obtained by representing spin operators $\sigma^x$, $\sigma^y$, and $\sigma^z$ on a given site in terms of four Majorana fermions $b^x$, $b^y$, $b^z$, and $c$ so that $\sigma^\alpha = i b^\alpha c$, Fig.~\ref{fig:intro}(a). The transformation from spin to fermions enlarges the Hilbert space. The physical subspace with spin algebra $[\sigma^\alpha, \sigma^\beta] = 2i \epsilon^{\alpha\beta\gamma} \sigma^\gamma$ has $D \equiv b^x b^y b^z c = +1$. The Hamiltonian for the Majorana fermions is quadratic in $c$ modes,
\begin{equation}
H = i\sum_{\langle mn \rangle}J_{\alpha_{mn}}u_{mn}c_m c_n,
\label{eq:Hmajorana1}
\end{equation}
where $\langle mn \rangle$ is a nearest-neighbor bond. 
$b$ modes are hidden in static $Z_2$ gauge variables $u_{mn} = -u_{nm} = i b_m^\alpha b_n^\alpha = \pm 1$. Plaquette variables $W$ are fluxes of the $Z_2$ gauge field: $W = (-u_{12}) u_{23} (-u_{34}) u_{45} (-u_{56}) u_{61}$, with a $-1$ factor for each link pointing from the odd sublattice to the even one. Operator $D_m$ implements a $Z_2$ gauge transformation on site $m$. Majorana fermions $b_m^\alpha$ and $c_m$ and link variables $u_{mn}$ are odd under $D_m$, whereas physical variables such as spins $\sigma_m^\alpha$ and $Z_2$ fluxes $W$ are even. Local constraints $D_m = +1$ express gauge invariance of physical states: $D_m |\psi\rangle = |\psi\rangle$.

In a static gauge background $\{u_{ij}\}$, the Hamiltonian (\ref{eq:Hmajorana1}) can be transformed to a diagonal form $H = \sum_{m} (\epsilon_m/2) (\psi^\dagger_m \psi_m - \psi_m \psi^\dagger_m)$, where $\psi^\dagger_m$ and $\psi_m$ create and destroy a physical fermion excitation with energy $\epsilon_m > 0$. The ground state is in the sector with $W=+1$ on all hexagonal plaquettes. We shall focus our attention on one of the gapped phases, where one of the coupling constants in Eq.~(\ref{eq:HKitaev}) dominates, e.g., $J_{z} > J_{x} + J_{y}$. We assume ferromagnetic couplings, $J_\alpha>0$, without loss of generality. The physics simplifies in the limit $J_{z} \gg J_x, \, J_y$, where low-energy states have parallel spins on strong ($z$) bonds. Fermion excitations $\psi$ are associated with breaking this alignment and have a high energy cost close to $2J_z$, so we shall refer to them as high-energy fermions. Low-energy excitations are vortices, $W = -1$, with  energy $J_x^2 J_y^2/8J_z^3$ \cite{Kitaev2006}. The effective Hamiltonian in this subspace turns out to be Kitaev's toric code \cite{Kitaev2003,Wen}, with effective spins $\tau_{mn}^z = \sigma_m^z = \sigma_n^z$ living on links of a rectangular lattice, Fig.~\ref{fig:intro}(b).  Crucially, vortex excitations come in two flavors---$e$ and $m$---depending on the plaquette. A honeycomb plaquette centered on a vertex (plaquette) of the toric-code lattice may host an $e$ ($m$) vortex. Thus $e$ and $m$ vortices live in alternating rows, Fig.~\ref{fig:intro}(b). 

Like the $e$ and $m$ particles of the toric code, honeycomb vortices of different flavors are mutual semions: winding an $e$ vortex around an $m$ vortex alters the sign of the wavefunction. By extension, a vortex pair $e\times m$ is a composite fermion: exchanging two such pairs gives a minus sign. A vortex pair $e\times m$ cannot be created out of the vacuum alone as that would violate conservation of fermion parity (fermions can only be created or destroyed in pairs \cite{Kitaev2001}). The application of an operator $\sigma_n^x$ or $\sigma_n^y$ creates an $e \times m$ vortex pair in adjacent rows, Fig.~\ref{fig:intro}(c), and misaligns spin $\sigma_n^z$ with its partner on a strong $z$ bond at a high energy cost of $2J_z$. In other words, a composite fermion $e\times m$ is created along a high-energy fermion $\psi$. Fermion parity is conserved but the process is effectively forbidden at low energies. Local conversion of the vortex flavor [Fig.~\ref{fig:intro}(d)] is forbidden for the same reason. 

\begin{figure}
\includegraphics[width=0.99\columnwidth]{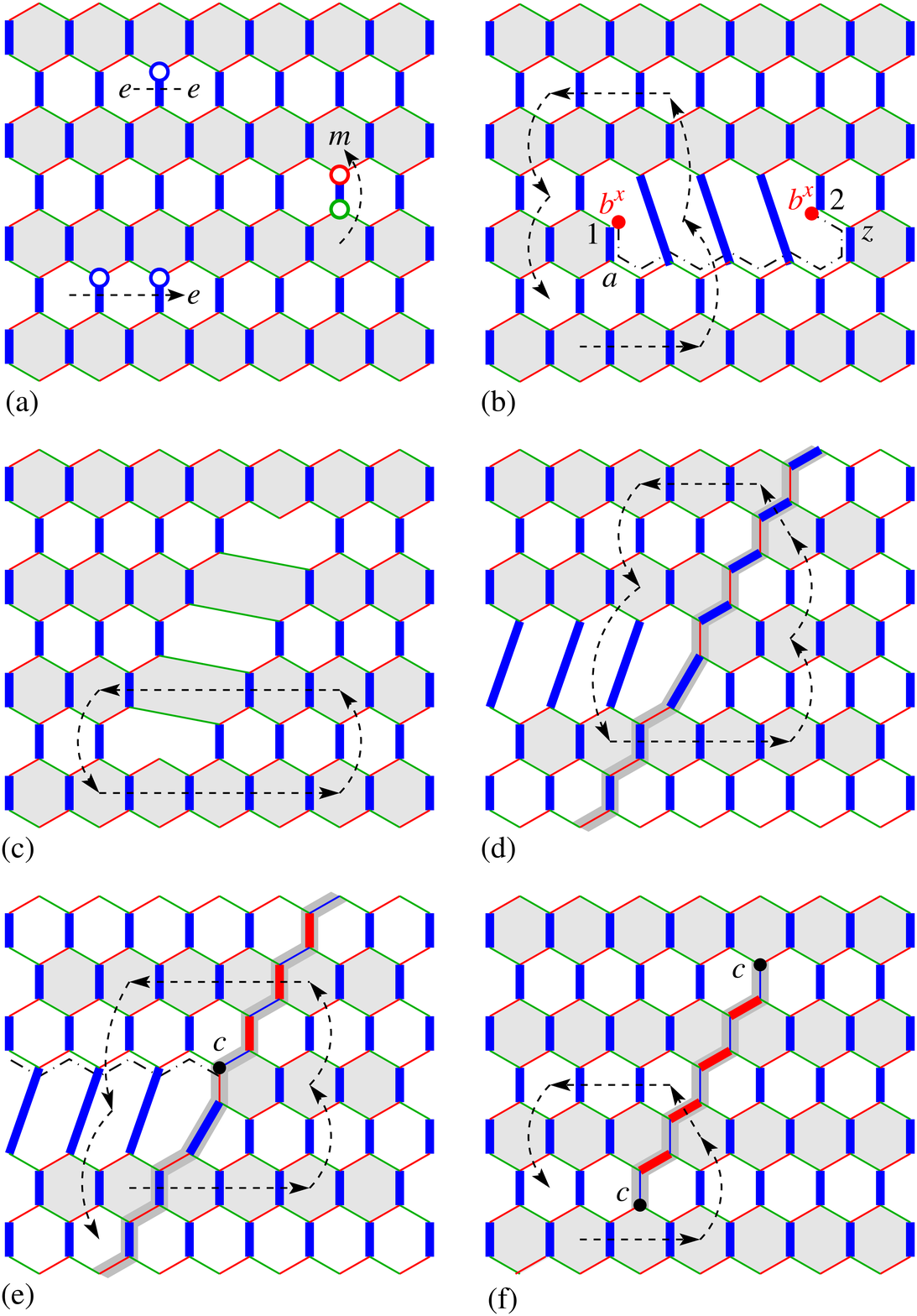}
\caption{(color) (a) Low-energy processes involving vortices. Open red, green, and blue circles denote application of $\sigma^x$, $\sigma^y$, and $\sigma^z$ operators. (b)--(c) A pair of dislocations with octagon cores: twists with $\mathbf B = \pm \mathbf a_1$ (b) and trivial with $\mathbf B = \pm \mathbf a_3$ (c). (d)--(e) A pair of dislocations with 5--7 cores and $\mathbf B = \pm \mathbf a_2$: trivial (d) and twists (e). (f) A string with altered dimerization with twists at the ends. Dashed lines are flux path. Dash-dotted lines are branch cuts. Filled circles are unpaired Majorana modes.}
\label{fig:fig2}
\end{figure}

Elementary low-energy processes are thus restricted to (a) creating and annihilating two vortices in the same row; (b) moving a vortex in its row; (c) moving a vortex to the next-nearest row, Fig.~\ref{fig:fig2}(a). The first two are accomplished by acting with an operator $\sigma_n^z$; the third by applying $\sigma_m^{x} \sigma_n^{y}$ or $\sigma_m^{y} \sigma_n^{x}$ on a strong bond $\langle mn \rangle$. 

Suppose two vortices of type $e$ are created out of the vacuum and then one of them completes a loop via the low-energy movements, returning to the starting point. In a lattice without defects, the vortex returns to the same row and can then be annihilated with its partner to bring the system back to its original ground state. However, if the loop encloses a dislocation, the vortex may return to an adjacent $m$ row and we end up with a composite $e\times m$ fermion. The apparent non-conservation of fermion parity is avoided if dislocations possess a Majorana mode $\beta$. Majorana modes of two dislocations can be combined to form a nonlocal fermion $\Psi = (\beta_1 + i \beta _2)/2$. A vortex winding around one of the dislocations alters the fermion number $\Psi^\dagger \Psi$ to compensate for the creation of the composite $e \times m$ fermion. 

In what follows we make an explicit construction of Majorana modes on dislocations in Kitaev's honeycomb model. A typical dislocation in graphene \cite{carpio2008dislocations,yazyev2010topological,cortijo2007effects} has a pentagon-heptagon (5--7) pair at its core. Although a 5--7 core preserves the threefold coordination of the honeycomb lattice, it alters the topology of link labels $x$, $y$, $z$, creating a string defect, Fig.~\ref{fig:fig2}(d). The topology can be preserved at the expense of removing a few bonds and creating sites with a reduced coordination number. The simplest such dislocation has an octagonal (8) core, Fig.~\ref{fig:fig2}(b) and (c). It can be synthesized in Kitaev's model by quenching a line of sites with a strong magnetic field \cite{You2,longerUNP}. We first discuss the more straightforward case of octagon dislocations. 

To act as a transformer of vortex flavor, a dislocation must have a Burgers vector $\mathbf B$ connecting plaquettes of different types, e.g., $\mathbf B = \pm \mathbf a_1$ or $\pm \mathbf a_2$ in the gapped phase with strong $z$ bonds. Fig.~\ref{fig:fig2}(b) shows a dislocation with $\mathbf B = \mathbf a_1$. It can be seen that a vortex winding around this dislocation via low-energy moves alters its flavor. Following Bombin \cite{Bombin, You1, Barkeshli}, we refer to such dislocations as \emph{twists}. Fig.~\ref{fig:fig2}(c) shows a dislocation with $\mathbf B = -\mathbf a_3$, which preserves the vortex type and is in this sense \emph{trivial}. 

\emph{Twist dislocations.} As can be seen in Fig.~\ref{fig:fig2}(b), the presence of a $\mathbf B = \mathbf a_1$ dislocation makes it impossible to partition the lattice into plaquettes of $e$ and $m$ flavors globally. Any locally consistent partition has a branch cut connecting two dislocation cores. An $e$ vortex crossing the branch cut turns into an $m$ vortex and vice versa.  

Because site 1 at the cusp of the octagonal core in Fig.~\ref{fig:fig2}(b) is missing a weak $x$ bond, its Majorana fermion $b_1^x$ is unpaired. To form a zero-energy (complex) fermion mode $\Psi$, we can combine $b_1^x$ with a dangling Majorana mode of another twist dislocation, e.g., $b_2^x$ in Fig.~\ref{fig:fig2}(b). The naive recipe, $\Psi = (b_1^x + i b_2^x)/2$, $\Psi^\dagger = (b_1^x - i b_2^x)/2$, does not work: the fermion parity $\pi_{12} = \Psi \Psi^\dagger - \Psi^\dagger \Psi = i b_1^x b_2^x$ is not a physical quantity because it is not gauge-invariant (odd under $D_1$ and $D_2$). This problem can be fixed by adding a gauge factor $U_{12} = (-u_{1a}) u_{ab} (-u_{bc}) \ldots u_{z2}$. We have
\begin{equation}
\Psi = (b_1^x + i U_{12} b_2^x)/2, 
\quad
\Psi^\dagger = (b_1^x - i U_{12} b_2^x)/2.
\end{equation}
The fermion parity $\pi_{12} = i U_{12} b_1^x b_2^x$ is now gauge invariant and can be expressed as a product of spin operators along the string $1ab \ldots z2$. Like a branch cut, a string does not have a well-defined position; only its ends are fixed at dislocation cores 1 and 2. 

We can now see that the state of this fermion mode is altered when a flux winds around either of the dislocations. When the path of the flux crosses the string $1ab \ldots z2$, the link variable $u_{mn}$ at their crossing changes sign. This alters the sign of $U_{12}$ changing this mode's parity. We have thus established that the variable $b_m^\alpha$ of an octagon cusp missing a weak bond $\alpha$ is the Majorana mode associated with a twist dislocation. Together with a Majorana fermion of another dislocation, it forms a zero-energy mode whose quantum state can be changed by winding a flux around one of the dislocations.  

Having established the nature of the Majorana modes at twist dislocations, we can estimate their tolerance to local perturbations. In the presence of a magnetic field $\mathbf h = (h,0,0)$, the dangling modes $b_1^x$ and $b_2^x$ are coupled to the rest of the system by the Zeeman term $- h \sigma_m^x = - i h b_m^x c_m$. This coupling lifts the degeneracy of the zero mode and induces its time evolution, an undesirable effect if the field is noise. We shall see that the splitting decays exponentially with the distance between dislocations. The adverse effects of local noise can be suppressed by keeping dislocations sufficiently far apart.

To compute the splitting of the zero mode, we integrate out the high-energy $c$ modes. Consider first a toy model of four Majorana modes: $c_1$ and $c_2$ strongly coupled to each other, $b_1$ and $b_2$ weakly coupled to them: $H = \frac{i E}{2} c_1 c_2 + \frac{i \lambda}{2}(b_1 c_1 + b_2 c_2)$. Integrating out the $c$ modes yields a low-energy Hamiltonian for the $b$ modes, $H_\mathrm{eff} = \frac{i \epsilon}{2} b_1 b_2$ with the splitting $\epsilon = \lambda^2/E$. Thus by integrating out the modes $c_1$ and $c_a$ on the strong bond $1a$ adjacent to the cusp, we generate the coupling $-i (h J_x/J_z) u_{1a} u_{ab} b_1^x c_b$ between the dangling Majorana $b_1^x$ and a more distant mode $c_b$. After repeating the process enough times, we generate an effective coupling between the dangling Majorana modes,  
\begin{equation}
H_\mathrm{eff} = \frac{2 h^2}{J_z}  
	\sum_\mathrm{paths} \frac{J_x^{n_x} J_y^{n_y}}{J_z^{n_x+n_y}}  
		i b_1^x u_{1a} u_{ab} \ldots u_{z2} b_2^x.
\label{eq:b-Heff}
\end{equation}
The sum is taken over paths $1ab\ldots z2$ with $n_\alpha$ links of type $\alpha$. Paths must alternate between weak and strong bonds and thus can propagate only upward or downward, staying within overlapping 60-degree wedges with vertices at the dislocations. This coupling only exists for dislocations on different sublattices. For $J_x/J_z = J_y/J_z = j \ll 1$, the energy splitting induced by the potential depends on the length $L$ of a path between dislocations as $j^{(L-1)/2}$. A similar interaction was found by \textcite{Willans} between vacancy-induced magnetic moments.

\emph{Trivial dislocations.} A trivial dislocation has a core of the same shape as its twist counterpart. However, thanks to a different orientation of the octagon core, the missing bond at its cusp is strong, Fig.~\ref{fig:fig2}(c). The missing bond leaves a dangling $b^z$ Majorana mode at the cusp. In addition, a trivial dislocation has a second free Majorana mode of the $c$ type. If the weak bonds are completely switched off, $J_x = J_y = 0$, the additional zero mode is the $c$ fermion at the cusp. At nonzero $J_x$ and $J_y$, but still in the gapped phase ($J_x + J_y < J_z$), the zero mode is a superposition of $c$ fermions in the vicinity of the cusp, as in the case of a vacancy \cite{Willans}. The two zero modes can be combined to form a local, gauge-invariant (and thus physical) degree of freedom that acts like a free magnetic moment. Its gyromagnetic tensor $g$ has only one nonzero component $g^{zz}$. 

A trivial dislocation thus behaves very much like a vacancy. Its unpaired Majorana mode $b^z$ is susceptible to local noise due to the presence of a second unpaired Majorana mode of the $c$ type that it could couple to. The additional mode is absent in a twist dislocation, so its unpaired Majorana mode is robust. 

\emph{5--7 dislocations.} Fig.~\ref{fig:fig2}(d) shows a 5--7 dislocation with a Burgers vector $\mathbf B = \mathbf a_2$. Even though it has the right Burgers vector, this dislocation is not a twist. The presence of two disclinations at the core changes the orientation of $x$ and $z$ bonds along a line extending from the core. A vortex crossing the defect line in low-energy motion comes off a Burgers contour and will not change its flavor upon winding around the dislocation. Plaquette types can be globally assigned without ambiguity using low-energy vortex motion. This dislocation is trivial and is thus need not host a free Majorana mode.  

The situation changes if the strength of exchange coupling is determined by the bond's orientation, rather than its type, Fig.~\ref{fig:fig2}(e). In this case, a vortex follows a Burgers contour and changes flavor, making the dislocation a twist. One of the sites at the dislocation core has a $c$ operator weakly coupled to its neighbors. The unpaired Majorana mode is a superposition of that $c$ operator with its neighbors along two 60-degree wedges extending in both vertical directions. Its interaction with other unpaired Majorana modes is similar to that of a $b$ mode at an octagon dislocation, Eq.~(\ref{eq:b-Heff}), with two distinctions: the $c$ mode couples in both vertical directions and does not require a magnetic field for coupling. This zero mode has a one-dimensional analog in Kitaev's Majorana chain with alternating weak and strong bonds \cite{Kitaev2001} and other fermionic models \cite{JackiwRebbi, SuSchriefferHeeger}. The shaded path in Fig.~\ref{fig:fig2}(d) contains a Majorana chain with regular alternation and a gapped excitation spectrum. The one in Fig.~\ref{fig:fig2}(e) has a defect---a domain wall between the two possible alternating patterns---that binds a zero mode.

The above examples demonstrate explicitly that in Kitaev's model dislocations acting as twists \cite{Bombin} are associated with unpaired Majorana modes. Whether a dislocation is a twist depends on its Burgers vector as well as on its internal structure. One lesson from this is that the mere presence of a dislocation is not sufficient for the existence of a single unpaired Majorana mode. Furthermore, that we were able to convert a 5--7 dislocation from trivial to a twist is a hint that perhaps a dislocation is not even necessary. Indeed, one could start with a lattice without dislocations and introduce a string defect, along which alternating weak and strong bonds are interchanged, Fig.~\ref{fig:fig2}(f). The ends of the string act as twists and possess free Majorana modes of the $c$ type. 

In his seminal paper on the honeycomb spin model, \textcite{Kitaev2006} posited the existence of unpaired Majorana modes associated with lattice dislocations. Such zero modes offer a realization of non-Abelian statistics: the degenerate ground state of the system can be transformed by winding fluxes around dislocations, which alters the states of both the fluxes and the unpaired Majorana modes. Because well-separated Majorana modes are insensitive to local perturbations, these transformations are robust against local noise and thus offer a path to fault-tolerant quantum memory. Kitaev's proposal rests on two observations: (1) the gapped topological phases of the model have two flavors of vortex excitations and (2) a dislocation may alter the vortex flavor upon braiding (defects termed twists by \textcite{Bombin}). We have studied several examples of dislocations in Kitaev's honeycomb model and found that they can act as twists under the right circumstances. We have identified the unpaired Majorana modes and characterized their sensitivity to local perturbations. We have also found that twists can be created by other types of topological defects such as the alternation of weak and strong bonds along a string. 

In a forthcoming paper \cite{longerUNP}, we will provide a detailed account of interactions between dislocations and introduce a diagrammatic method of calculating the vortex energy. We will also discuss related issues such as the effect of additional dislocation pairs on the topological degeneracy of the system and the possibility of creating synthetic dislocations via the application of a magnetic field along a line of sites. If the gapped phase of the honeycomb model is physically realized in the future,  the latter will provide an implementation of a quantum memory device.

\emph{Acknowledgments.} The authors thank L. Balents, J.T. Chalker, R. Moessner, M. Oshikawa, S. Parameswaran, Y. Wan, X.-G. Wen, and H. Yao for useful discussions. They acknowledge hospitality of the Kavli Institute for Theoretical Physics and of the Aspen Center for Physics, where part of this work was done. This work was supported in part by the US Department of Energy, Office of Basic Energy Sciences, Division of Materials Sciences and Engineering under Grant No. DE-FG02-08ER46544 (JHU), by the US National Science Foundation under Grants No. PHY-1066293 (ACP) and PHY11-25915 (KITP), by Fondecyt under Grant No. 11121397 and Conicyt under Grant No. 79112004 (AIU).

\bibliography{dislocations}

\begin{thebibliography}{16}
\bibitem[{foo()}]{footnote}
\bibinfo{note}{We disregard other possible sources of degeneracy such as the
  topology of the system.}
\end{thebibliography}

\end{document}